# STRONGLY NONLINEAR WAVE DYNAMICS IN A CHAIN OF POLYMER COATED BEADS

C. Daraio[1], V. F. Nesterenko[1,2*]

[1]Materials Science and Engineering Program, University of California at San Diego, La Jolla CA 92093-0418 USA

[2]Mechanical and Aerospace Engineering Department, University of California at San Diego, La Jolla CA 92093-0411 USA

**Abstract.** Strongly nonlinear phononic crystals were assembled from a chain of Parylene-C coated steel spheres in a polytetrafluoroethylene (PTFE) holder. This system exhibits strongly nonlinear properties and extends the range of materials supporting "sonic vacuum"-type behavior. The combination of a high density core and a soft (low elastic modulus) coating ensures a relatively low velocity of wave propagation. The beads contact interaction caused by the deformation of the Parylene coating can be described by classical nonlinear elastic Hertz theory despite the viscoelastic nature of the polymer and the high strain rate deformation of the contact area. Strongly nonlinear solitary waves excited by impacts were investigated experimentally and compared to chains composed of uniform steel beads. Fracture of the polymer coating was detected under relatively large pulse amplitude.

**Introduction**

In the recent years the interest in understanding the dynamic behavior of granular materials has known a substantial increase [1-39]. Strongly nonlinear one dimensional chains of different materials represent the simplest examples of the media important for the development of a new wave dynamics.

The strongly nonlinear behavior of granular matter arises from its inherent "double nonlinearity" due to the Hertzian contact interaction between granules under compression and the zero interaction force under tensile stresses. This non-classical, strongly nonlinear behavior appears only if the granular material is "weakly" compressed ("sonic vacuum") [1-6,18]. The particles contact interaction under "strong" precompression alters the wave propagation behavior from the strongly nonlinear type to the linear or weakly nonlinear (KdV-type) [12]. A unique property deriving from this "tunability" of the contact interaction is the ability to control the wave propagation regime simply adjusting the amount of precompression acting on the system [38] or varying the elastic properties at the particles contacts or mass of particles.

Coating a high density particle such as the steel beads considered in this study with a soft polymeric layer might ensure a better control of impulse transformation and propagation characteristics allowing applicability in a large variety of engineering solutions. The results obtained from these analyses may lead to the discovery of novel applications (i.e. sound focusing devices (acoustic lenses), very short transmission delay lines, sound and shock absorption layers and sound scramblers). Granular beds composed from iron particles are used for designing systems able to transform impacts from heavy strikers and contact explosion [18]. The soft coating is important for delaying and reducing the amplitude and wave speed in the medium adjusting, for example, the protective granular barrier to a wider spectrum of dynamic loads. Parylene-C is a very versatile visco-elastic material, widely biocompatible, with a very low friction coefficient, and low elastic modulus. Its properties, combined with the wide commercial availability of Parylene-C thin film coating, allowed it to be suitable for the assembling of strongly nonlinear systems. To the best of our knowledge no other results on the pulse propagation in chains composed of coated particles were previously published.

**Theoretical Analysis**

For the theoretical treatment, we consider that the core of the particles behaves as a rigid body and the depth of the deformed contact zone is smaller than the thickness of the coated layer. In the initial conditions the beads have all point contacts with their neighboring particles and the particles interaction is limited only to the deformation of their coating. Finally we assume that the time of the particle interaction is significantly longer compared to the time of the wave propagation inside the particle ensuring the application of a quasistatic interaction law. Under these assumptions using the Hertzian expression of force between the granules [40],

$$F = \frac{2E}{3(1-v^2)} \left(\frac{R'+\Delta}{2}\right)^{1/2} \left(2(R'+\Delta)-(x_2-x_1)\right)^{3/2} \quad (1)$$

we obtain the following equation of motion for a one dimensional chain of identical uniform coated beads (Fig. 1) similar to the one presented in [18]:

$$\ddot{u}_i = A(\delta_0 - u_i + u_{i-1})^{3/2} - A(\delta_0 - u_{i+1} + u_i)^{3/2}, \qquad (2)$$

where

$$A = \frac{E\, a^{1/2}}{3(1-\nu^2)M}, \quad a = 2(R'+\Delta)$$

and

$$M = m_{ss} + m_p = \frac{4}{3}\pi R'^3 \rho_{ss} + \frac{4}{3}\pi\left((R'+\Delta)^3 - R'^3\right)\rho_p.$$

Here $\delta_0$ is the initial displacement between particles due to the static precompression, $E$ is the elastic modulus and $\nu$ is the Poisson's coefficient of Parylene-C, $\Delta$ represents the Parylene coating thickness. The diameter of the particles is $a=2R$, the mass $M$ and the displacement of the $i$-th particle is $u_i$.

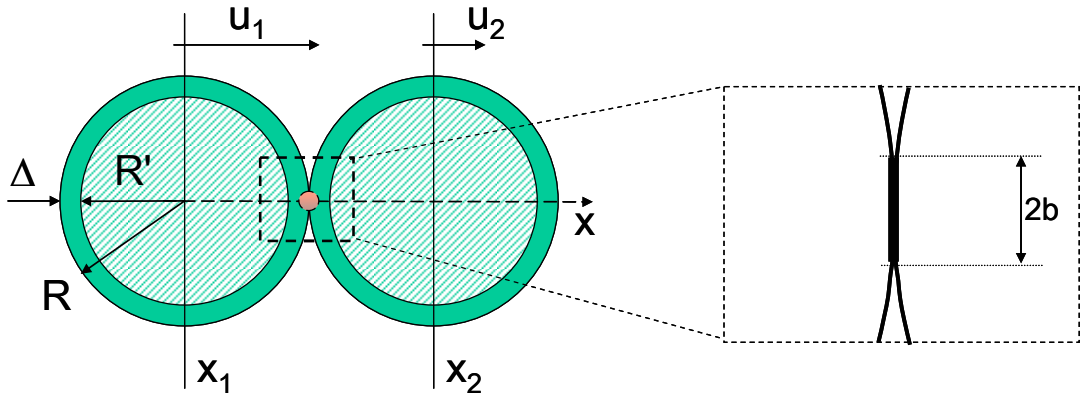

FIG. 1. Schematic diagram showing the contact interaction area between two Parylene coated particles. The inset shows an enlargement of the contact interaction area.

When a weak precompression is acting on the system the long wave equation for the particle displacement u is:

$$u_{tt} = -c^2 \left\{ (-u_x)^{3/2} + \frac{a^2}{10} \left[ (-u_x)^{1/4} \left( (-u_x)^{5/4} \right)_{xx} \right] \right\}_x \tag{3}$$

where $-u_x > 0$, $c^2 = \dfrac{E\, a^3}{3(1-v^2)M}$ and $c_0 = \left(\dfrac{3}{2}\right)^{1/2} c \xi_0^{1/4}$.

Here $\xi_0$ is the initial strain in the system. It should be noticed that $c$ does not represent the sound speed in the chain, while $c_0$ corresponds to the long wave sound speed at initial strain $\xi_0$. This equation for high amplitude pulses (or for negligible precompression) has no characteristic wave speed. In this case, despite its complex nature, Eq. (2) has simple stationary solutions supporting supersonic solitary waves with finite length equal to only five particle diameters [1-4,18].

In the leading approximation, the force $F_m$ is related to the maximum strain $\xi_m$ in the solitary wave as follows:

$$F_m = \frac{4E}{3(1-v^2)}(R' + \Delta)^2 \xi_m^{3/2} . \tag{4}$$

Also, from [18] we know that for a solitary wave in "sonic vacuum" the speed is equal:

$$V_s = \frac{2}{\sqrt{5}} c \xi_m^{1/4} . \tag{5}$$

Combining Eq. (4) and Eq. (5) we obtain the relation between the solitary wave speed and the maximum force in the solitary wave propagating in the chain of composite particles in "sonic vacuum":

$$V_s = \frac{2}{\sqrt{5}} \left( \frac{E\, a^{7/2}}{3(1-v^2)} \right)^{1/3} \frac{F_m^{1/6}}{\sqrt{M}} . \tag{6}$$

It's evident that the speed of this wave has still a nonlinear dependence on the maximum strain $\xi_m$ or force $F_m$ (particle velocity $\upsilon_m$) and can be infinitesimally small if the amplitude of the wave is small. This phenomenon is of fundamental interest because Eq. 3 is more general than the well known weakly nonlinear KdV equation. The existence of this unique wave was verified in numerical calculations and in experiments for chains composed of uniform beads of different materials [18].

The addition of significant initial precompression on the system is known to modify the wave propagation behavior, for example due to the presence of gravitational precompression in the vertical alignment of the chains or due to the added prestress by an other external source [38]. The speed of the solitary wave $V_s$ in a chain of coated beads with applied static precompression ($F_0$) is derived in a similar way as in [33] for uniform particles:

$$V_s = \left(\frac{2}{15}\right)^{1/2} \left(\frac{3E^2 F_0 a^7}{M^3(1-\nu^2)^2}\right)^{1/6} \frac{1}{f_r^{2/3}-1} \left(3 + 2f_r^{5/3} - 5f_r^{2/3}\right)^{1/2}. \tag{7}$$

The speed $V_s$ has a nonlinear dependence on the maximum force $f_r = F_m/F_0$ but can be considered approximately constant at any relatively narrow interval of its amplitude due to the small exponent present in the power law. Equation (6) can be obtained as a limit of Eq. (7) when the precompression ($F_0$) is approaching zero.

**Experiments and discussion**

We assembled strongly nonlinear chains from steel bearings balls (diameter of 4.76 mm and 2.48 mm) coated with a 50 μm thick layer of Parylene-C polymer (AcraBall

Inc.) via vaporization processes (Fig. 2(a)). The properties of Parylene-C provided by the manufacturer are the following: Young modulus, $E$ = 2760 MPa, density = 1289 Kg/m$^3$, Poisson's coefficient $v$ = 0.388 [41]. The yield strength ($Y$) of thin films of Parylene-C is reported to be at 59 MPa [42].

Parlylene-C is a polymeric viscoelastic material with a low elastic modulus. It is widely used in different areas from electronic to biomedical applications [42]. Its properties can be very attractive to ensure a low speed of soliton propagation and a good tunability of the system. At the same time it is not evident that a chain formed from this type of composite beads would support strongly nonlinear solitary waves based on Hertzian interaction, as for example chains made from uniform linear elastic materials like steel [2,4,10,14,18,28,37]. In particular, the role of dissipation and deviation from the linear elastic law under high strain rate deformation (about $10^2 - 10^3$ s$^{-1}$) can influence the contact interaction and the wave dynamics. Additionally, the interfacial phenomena between the Parylene coating and the steel core, for example due to the weak adhesion, may contribute to the contact deformation and dissipation and affect the dynamic behavior of the chains. Chains of beads made from Nylon [14] with elastic modulus ~23% larger than Parylene-C and chains made of PTFE [33] with a nominal elastic modulus ~5 times smaller than Parylene-C demonstrated the validity of the Hertz type interaction law and supported propagation of strongly nonlinear solitary waves.

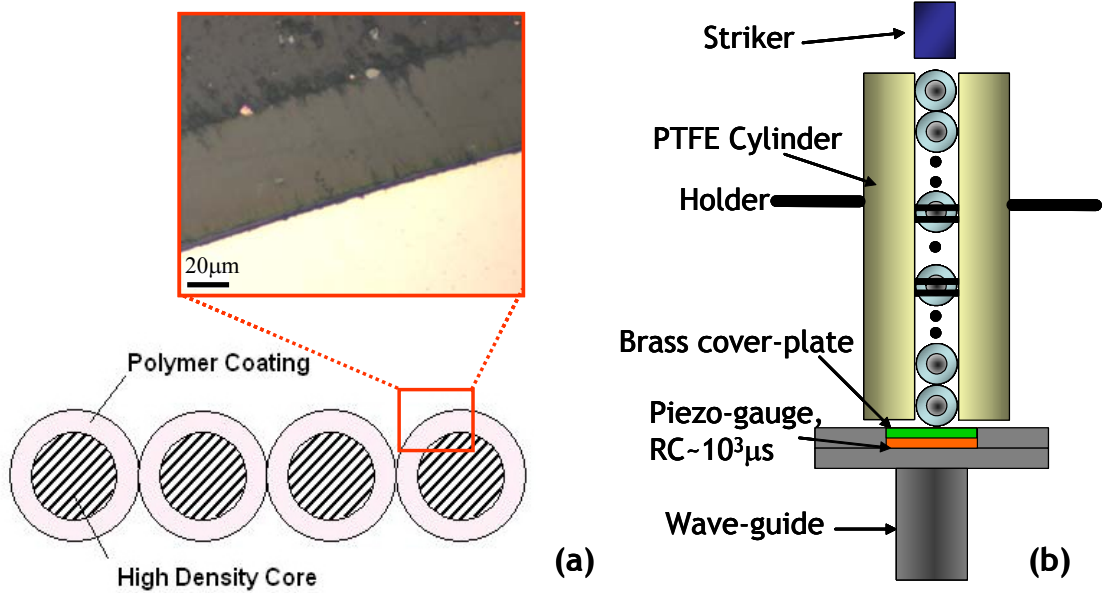

FIG. 2. Experimental set-up for testing of 1-D strongly nonlinear phononic materials. (a) schematic diagram of the chain and optical micrograph showing the Parylene-C polymer coating (b) set up used for the single chain testing, showing the particles with embedded sensors.

One dimensional phononic crystals were assembled from 21 Parylene-C coated steel beads (with diameter a=4.86 mm and mass 0.44 g) vertically aligned in a PTFE (polytetrafluoroethylene) holder with inner diameter 5 mm (Fig. 2(b)). Pulses of different durations and amplitudes in the 1-D phononic crystals were generated by impact of an alumina ($Al_2O_3$) cylinder (0.47 g and 1.23g) on the top particle of the chain from different heights.

In the described set-up we included three calibrated piezo-sensors (RC ~$10^3$ μs) connected to a Tektronix oscilloscope similar to [33,38,39]. Two lead zirconate titanate based piezo-gauges with Nickel electrodes and custom microminiature wiring (Piezo Systems, Inc.) were embedded inside two of the Parylene-C coated particles. The original beads were cut into two parts and the wired piezo-elements were glued between the steel parts. The total mass of the assembled beads with sensors was approximately equal to the mass of the original particle (0.44 g) to preserve the uniformity of the chain. This design ensured a precise calculation of the speed of signal propagation simultaneously with the measurement of the forces acting in the particles. A third piezo-gauge (Kinetic Ceramics, Inc.) was bonded with epoxy on the electrode pads for contacts and reinforced by a 1 mm brass plate on the top surface. This wall-sensor was then placed on the top surface of a long wave guide (vertical steel rod) ending into a steel block to avoid possible reverberations in the system. It was calibrated using linear momentum conservation law during single impact by a steel ball. Sensors embedded in the particles were calibrated by comparison with the signal from the sensor at the wall under controlled impact loading.

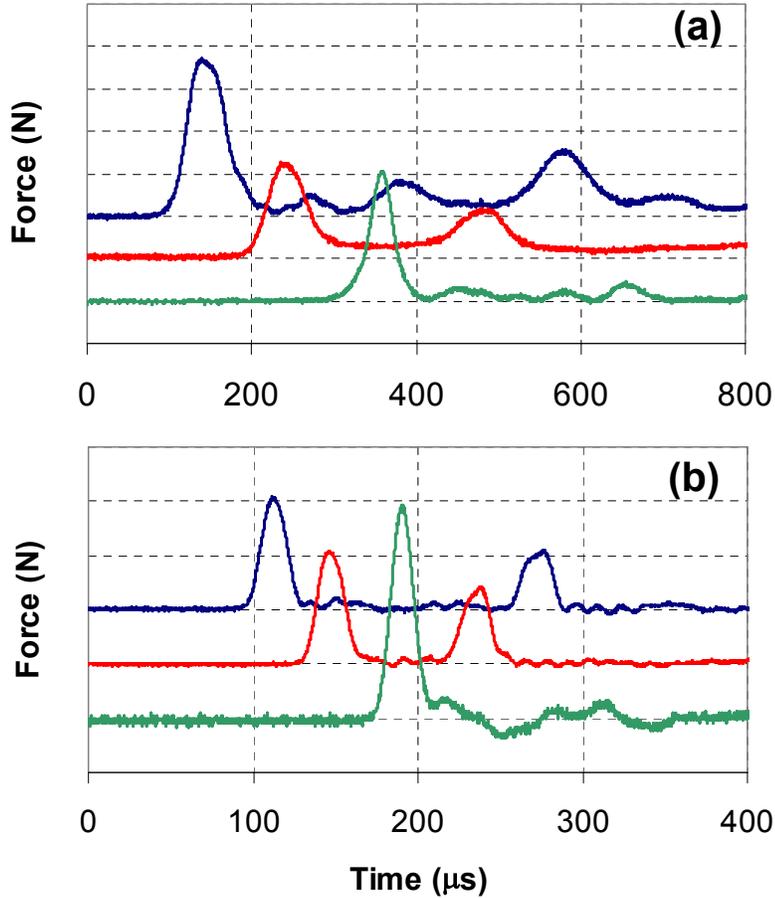

FIG. 3. Pulses generated in chains composed of (a) Parylene-C coated steel balls (4.86 mm diameter) and of (b) bare stainless steel balls (4.76 mm diameter) for comparison of their amplitudes, speed, width and attenuation. In both chains sensors are placed in the 9$^{th}$ ball from the bottom (top curves), 5$^{th}$ ball from the bottom (middle curves) and at the bottom wall. The velocity of striker (0.47 g $Al_2O_3$ cylinder) was 0.44 m/s. The scale of the vertical axis is 1 N for (a) and 5 N for (b).

The dynamic behavior of the chains composed of Parylene-C coated beads and steel beads was initially tested (Fig. 3) in conditions usually resulting in exciting single solitary waves. This is observed when the mass of the striker is close to the mass of one bead [4,18,33,38,39]. The speed of signal propagation in the chains was calculated dividing the distance between the sensors by the peak to peak time interval. The accuracy of the speed measurement can be estimated about 10 % due to the uncertainty in the sensors alignment (±1 mm for each sensor).

The experimental results corresponding to a chain composed of 21 Parylene-coated beads (4.86 mm diameter, 0.44 g mass) are shown in Fig. 3(a). It is apparent that under these conditions of loading a single pulse is mainly excited. In comparison to the bare steel beads case (Fig. 3(b)) the composite particles showed some additional small pulses, following the main pulse, especially noticeable in the first (uppermost) gauge record. Probably this behavior is related to the coating on the beads composing the system. The speed of the main pulse propagating in the Parylene coated beads chain (Fig. 3(a)) is 197 m/s, being significantly below the level of sound speed in gases at normal conditions. The main pulse duration was about 130 μs with width equal ~5.3 particle diameters being close to the length predicted in the theory based on the stationary solution of Eq. (3) based on Hertz interaction law [1,18]. The strain rate in the deformed coating was in the interval $10^2$-$10^3$ s$^{-1}$. This allows us to conclude that the Parylene-C coated beads chain supports the propagation of strongly nonlinear solitary waves, despite the viscoelastic nature of the polymer in the coating layer. It's worth noticing that the amplitude of the solitary waves traveling through the chains (especially in the Parylene-C based system) might be heavily affected by the presence of dissipation. Its effects will be addressed in our future research.

It is interesting to compare the behavior of the polymer coated beads with a chain of bare steel beads (Fig 3(b)) of similar diameter (equal to the diameter of the steel core of the coated beads) and under the same type of loading (same impact striker and same velocity). Steel beads are well known to support strongly nonlinear behavior [2,4,10,14,18,28,37]. It is evident from comparison between Fig. 3(a) and 3(b) that the behavior of the coated and the not coated beads differs dramatically in the speed of the propagating pulse (578 m/s for steel beads versus 197 m/s for composite beads). Although, both chains are forming a solitary pulse with width close to 5 particle diameters which is similar to a size of a stationary solution of Eq. (3) [1,4,18]. Reducing the amplitude of the signal traveling through the system to ~0.25 N we were able to achieve solitary waves speeds in the chain composed of Parylene coated beads down to ~130 m/s, which is more than two times smaller than the smallest speed of a solitary wave detected for solid nylon beads [14] and it is comparable with the one obtained with solid PTFE particles [33]. In principle a "sonic vacuum" type media can be designed to support solitary waves with indefinitely small amplitude and speed of propagation. It is reasonable to expect detectable solitary waves with a speed within the order of magnitude of 10 m/s or lower, useful for example in acoustic delay lines and shock pulse mitigation devices.

It is interesting to follow the dependence of the solitary wave speed on amplitude and compare it with the results from long wave approximation. This agreement was previously demonstrated for strongly nonlinear chains of beads made from different materials [2,4,10,14,18,28,33,37].

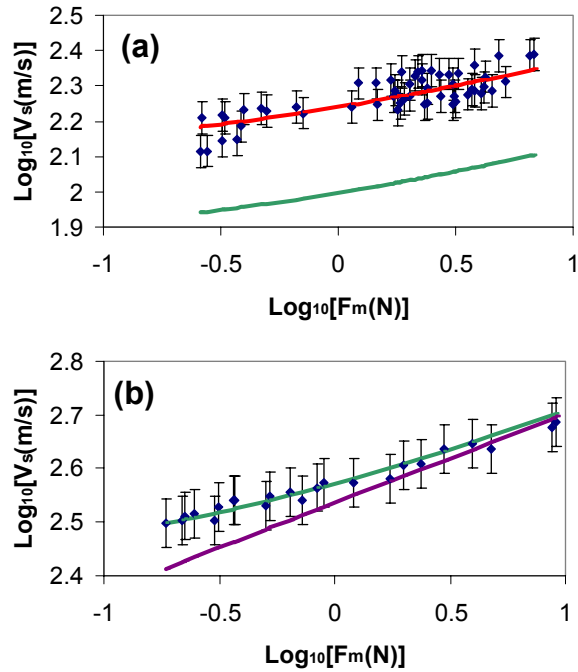

FIG. 4. Dependence of the wave speed on the amplitude of the solitary wave in the chains assembled from coated spheres (a) and from stainless steel beads (b). (a) Experimental vales (solid dots) and theoretical curves corresponding to the long wave approximation including gravitational preload. The lower curve corresponds to a value on elastic modulus E=2760 MPa, the upper curve matching the experimental values corresponds to E=15 GPa. (b) Experimental values obtained in a uniform stainless steel chain with the long wave approximation at E=193GPa for a "sonic vacuum" (lower curve) and for a gravitationally precompressed chain (upper curve).

To compare the experimentally measured speeds of the solitary waves at different amplitudes with the theoretical prediction derived from the long wave approximation we ran a series of experiments at different striker velocities. The results are summarized in Fig. 4(a) for the parylene coated beads. Here, the curves plotted represent the values obtained from the long wave approximation for a gravitationally loaded chain (Eq. (7)) for two different values of elastic moduli.

The lower curve in Fig. 4(a) corresponds to the value of elastic modulus for Parylene-C provided by the manufacturer ($E$=2760 MPa) [40]. The experimental values were matched by the theoretical curve (Eq. (7)) with the value of the elastic modulus $E$=15 GPa (see Fig. 4(a), upper curve). It should be mentioned that elastic moduli of polymers are very sensitive to the conditions of testing. For example, measurements of the elastic modulus of Parylene-C based on membrane deflection measurements gave a significantly higher value of 4.75 GPa [41] than one provided by manufacturer. It is evident that in the polymer based strongly nonlinear system there is a large discrepancy between the experimental pulse's speed values and the theoretical curve derived assuming the value of the standard elastic modulus $E$=2760 MPa. In Fig. 4(a) the value of the elastic modulus matching the experimental data is ~5.4 times higher than its value provided by the manufacturer. A similar difference was observed also for another polymer based granular material in [33] where the elastic modulus of PTFE appeared to be ~3 times higher than its nominal value (and matched the Young's modulus extrapolated from the Hugoniot measurements [42]). It is likely that in polymer-based strongly nonlinear systems the elastic modulus might be stress and strain dependent [43]. Further research on the value of elastic modulus of Parylene-C coating under conditions of dynamic deformation and validity of Hertz approximation for contact law is necessary to clarify the observed phenomenon.

Results for the pulse traveling in the bare stainless steel beads chain are presented in Fig. 4(b). In this case the curves represent the long wave approximation at $E$=193 GPa for steel in the "sonic vacuum" (lower curve) and in the gravitationally precompressed case (upper curve) agreeing with experiments in [2,4,10,14,18,28,37].

In both cases, the theoretical estimate appears to be strikingly close to the experimental results, especially if taking into account that no adjustable parameter was used in the theoretical consideration.

We may conclude that chains of polymer coated beads with a high density core support the propagation of a strongly nonlinear solitary wave, being another realization of a "sonic vacuum" type phononic crystal with an exceptionally low speed of signal as observed in chains of beads made from other polymeric materials like Nylon or PTFE [14,33].

Another remarkable feature of "sonic vacuum" type systems without dissipation is the very fast decomposition of a relatively long initial pulse into a train of solitary waves [4,18]. This type of pulse can be created using a striker with a mass higher than the mass of the particles in the chain (0.44 g). Apparently this property may be obscured by the strong dissipation in the polymer based viscoelastic systems. We used a striker with a mass $m_s$=1.23 g. The number of solitary waves with significant amplitude in which the longer initial pulse is expected to decompose throughout the chain is comparable to the ratio of the striker mass and the mass of the beads in the chain (~3 in this particular experimental set-up) [1,3,4,6]. The results of this experiment for a Parylene coated beads chain and for a chain made from bare stainless steel beads are shown in Fig. 5.

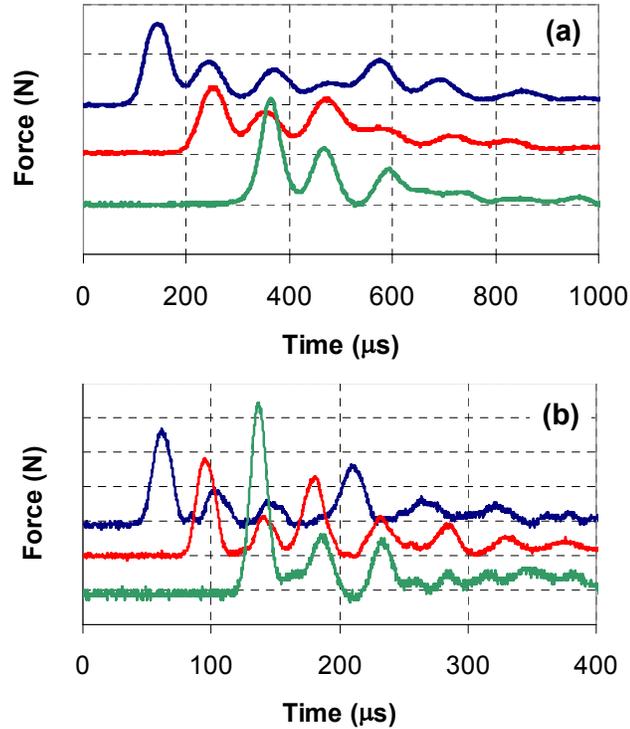

FIG. 5. Train of solitary waves generated in a 21 particles chain composed of (a) Parylene-C coated steel balls (4.86 mm diameter) chain and (b) bare stainless steel beads (4.76 mm diameter) chain for comparison. Sensors are placed in the 9$^{th}$ ball from the bottom (top curve), 5$^{th}$ ball from the bottom (middle curve) and at the bottom wall. The velocity of striker (1.23 g Al$_2$O$_3$ cylinder) was 0.44 m/s. Y-axis scale is 1 N for (a) and 5 N for (b).

It is evident that the polymer-based strongly nonlinear chain demonstrates a clear tendency to quickly decompose the initial impulse in a train of 3 solitary waves on a distance comparable with the soliton width (Fig. 5(a)) as it is also the case for the much less dissipative stainless steel based system (Fig. 5(b)). In both cases the impulse appears to be almost completely split at the wall, after traveling through only 21 particles, despite the significant difference in pulse speed. Previous experimental work with chains of steel, glass, brass, nylon and PTFE beads [1,2,4,10,14,18,33] validated the prediction of the existence of strongly nonlinear solitary waves as stationary solutions of the wave equation (Eq. (3)). At the same time, it is evident that the bare stainless steel chain supports a more pronounces splitting of the initial pulses into a train of solitary waves. This is most probably related to the higher dissipation in the contact area between the coated particles. The presence of an increased dissipation in the coated beads is evident from the faster decay of the pulses and the larger overlap of the peaks in the train. The presence of an increased viscous dissipation might contribute to a delay in the signal splitting, as observed in [44] for a chain of particles immersed in liquids of different viscosities (oil and glycerol).

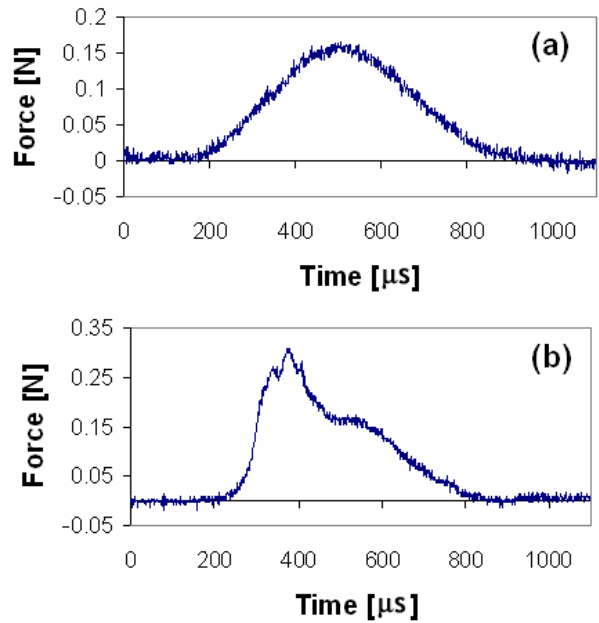

FIG. 6. Pulse generated by a **s**ingle solitary wave (a) and the shock like nonstationary pulse (b) detected on the wall supporting a chain of 22 smaller ParyleneC balls (2.48 mm diameter, m=0.04 g). (a) Impact at velocity 0.89 m/s with a 2 mm diameter steel ball

(0.036 g); (b) Impact at velocity 0.89 m/s with a 4.76 mm diameter PTFE ball with larger mass 0.123 g.

It is important to investigate the influence of particle size on the system's behavior for application purposes. Reducing the diameter of the Parylene-C beads may result in different stresses and strain rate conditions at the contact area affecting the behavior of the system specifically increasing the dissipation, as it was observed in a chain of PTFE beads [33]. We conducted experiments with smaller diameter steel balls (2.38 mm) coated by a 50 μm Parylene-C layer (total bead diameter is 2.48 mm). Experimental results are presented in Fig. 6. The sensor in this particular set-up was placed only at the wall for simplicity. Impacts were generated using a 2 mm diameter steel ball (0.036 g, Fig 6(a)) and a 4.76 mm diameter PTFE ball (0.123 g, Fig 5(b)) at velocity 0.89 m/s with no pre-compression acting on the chains (other than the practically negligible gravitational loading due to vertical set-up). It appears that the smaller diameter Parylene-C coated particles tend to form a single solitary wave when impacted by a striker of the same mass as a bead. This is a typical feature of the "sonic vacuum"-type behavior (Fig. 6(a)) [18]. When impacted by a striker with higher mass the effect of dissipation appears to delay the solitary wave splitting (Fig. 6(b)).

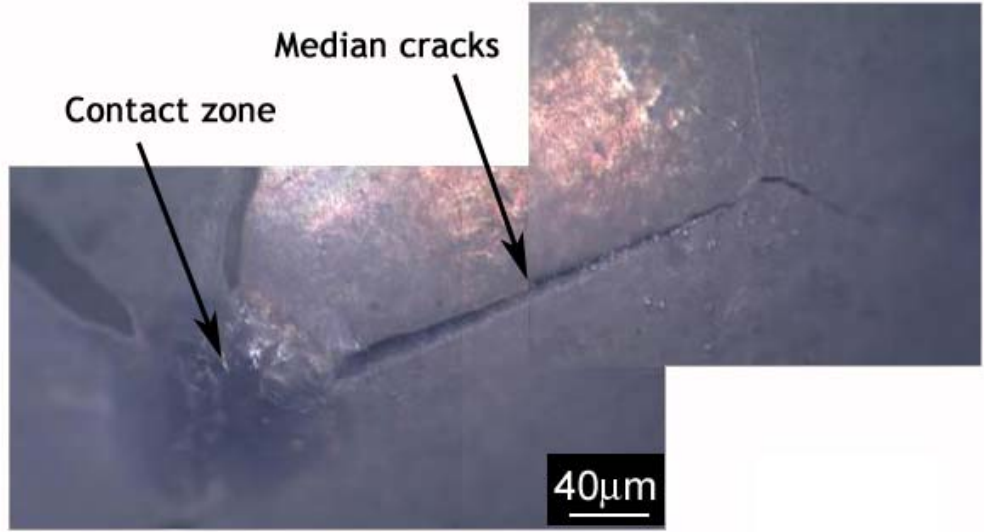

FIG. 7. Typical four median cracks originating away from the deformed contact zone observed after high amplitude dynamic testing with a maximum force on the level of ~3 N. Note that they appear before the Hertzian ring cracks which are supposed to originate on the outside area of the contact zone.

Parylene can be plastically deformed even at room temperature [41]; this restricts the amplitude of the pulses that can be treated based on the elastic approach. Cracking of the coating layer was seldom observed at higher dynamic force amplitude ranges (≥3 N). Despite this, we observed a consistent behavior in terms of the wave speed dependence on amplitude (Fig. 4(a)). Typical cracks morphology is depicted in Fig. 7. The development of median cracks before the Hertzian cracks is evident. We explain this phenomenon in analogy to the behavior of polymers under spherical indentation [45]. Finite element studies demonstrated that under significant displacements in the contact area the polymeric material is pushed outward causing hoop tensile stresses just outside the contact zone. In our experiments, the median cracks appeared prior to any ring cracks around the contact area, in agreement with the numerical results in [45]. We may expect that under very large contact forces and high strains in the contact area, the steel core might come into direct contact increasing the stiffness of the system therefore increasing the speed of signal propagation.

**Conclusion**

Chains composed of steel beads of different diameters coated by a 50 μm layer of Parylene-C were assembled and investigated under different conditions of impact loading. It was demonstrated for the first time that chains of composite particles with a "hard" core and a "soft" interacting contact do support the Hertzian type of interaction and the formation of single and trains of strongly nonlinear solitary waves. The "soft" interacting coating enabled a low signal propagation speed due to its exceptionally low elastic modulus and supported the formation of strongly nonlinear solitary waves despite its viscoelastic nature. The formed solitary waves broke the "sound barrier" having a speed of propagation well below sound speed in air. Decomposition of the incoming signal into trains of solitary waves was observed only for the larger diameter beads. At relatively high amplitudes of solitary waves we observed the formation of median cracks prior to the formation of ring Hertzian cracks.

**Acknowledgements**

This work was supported by NSF (Grant No. DCMS03013220).